\documentclass[%
reprint,
amsmath,amssymb,
aps,
prl,
]{revtex4-2}

\usepackage{graphicx}
\usepackage{bm}
\usepackage{hyperref}

\begin{document}

\title{Magnetic phase diagram of the three-dimensional doped Hubbard model}

\author{Liam Rampon$^{1,2}$}
\email{liam.rampon@polytechnique.edu}
\author{Fedor \v{S}imkovic IV$^{1,2}$}
\author{Michel Ferrero$^{1,2}$}

\affiliation{
$^1$CPHT, CNRS, Ecole Polytechnique, Institut Polytechnique de Paris, 91128 Palaiseau, France\\
$^2$Coll\`ege de France, 11 place Marcelin Berthelot, 75005 Paris, France}

\date{\today}

\begin{abstract}
  We establish the phase diagram of the Hubbard model on a cubic lattice for a wide range of temperatures, dopings and interaction strengths, considering both commensurate and incommensurate magnetic orders. We use the dynamical mean-field theory together with an efficient method to compute the free energy which enable the determination of the correct ordering vectors. Besides an antiferromagnetic state close to half-filling, we identify a number of different magnetic spiral phases with ordering vectors $(q,\pi,\pi)$, $(q,q,\pi)$ and $(q,q,q)$ as well as a region with close competition between them, hinting at spatial phase separation or at the onset of a stripe phase. Additionally, we extensively study several thermodynamic properties with direct relevance to cold-atom experiments: the entropy, energy and double-occupancy.
\end{abstract}

\maketitle

\emph{Introduction ---}
Strongly correlated electronic systems are particularly interesting due to the plethora of many-body physical effects they exhibit, ranging from magnetic phases, superconductivity to metal-to-insulator transitions and spatial phase separation. Generally, such systems are hard to solve controllably in most interesting regimes due to the exponentially growing Hilbert space and the various associated manifestations of the infamous fermionic sign problem.

The paradigmatic Fermi-Hubbard model \cite{Hubbard1963, Hubbard1964, Kanamori1963, Gutzwiller1963, Hubbard_gen_1, Hubbard_gen_2} is considered to be the minimal representative example of this class of systems, which, despite its apparent simplicity, manages to harbor most of the aforementioned physical effects.  Simultaneously, it eludes controlled numerical treatment, as it is exactly solvable only in the limiting cases of one and infinite dimensions. The two dimensional variant of the Hubbard model is the most extensively studied and most famous due to the relation to high-temperature superconductivity in cuprates \cite{Hubbard_gen_1} and more recently nickelates and paladates \cite{kitatani_palladates}, which can all be approximately described as two-dimensional layers weakly coupled in the third dimension. However, many strongly correlated materials such as i.e. perovskites \cite{klebel_knobloch}, are truly three-dimensional, with a hopping integral in the third dimension which can be comparable to the first two.

At the same time, the Hubbard model is by itself of experimental interest, since it has been simulated in various dimensions via ultra-cold atoms loaded in optical lattices \cite{Cold_atoms_AFM_1, Cold_atoms_AFM_2, Cold_atoms_AFM_3, Cold_atoms_AFM_4, zwierlein_spin_correl, bakr_canted_AF, bloch_magnetic_polaron, kohl_bilayer, bloch_chains_haldane_phase} . Despite remarkable recent improvements in these experiments, they are still limited to moderately high temperatures and can largely benefit from theoretical predictions for the currently inaccessible low-temperature regimes.

In two dimensions, the Mermin-Wagner theorem states that long range magnetic order is destroyed at finite temperatures by thermal fluctuations. This has been carefully documented by various numerical studies, mainly at half-filling (one electron per lattice site) where the sign problem is suppressed due to particle-hole symmetry. Upon decreasing the temperature, the model undergoes a crossover from the paramagnetic phase to a quasi-ordered regime with antiferromagnetic fluctuations and the system is only truly ordered at zero temperature ~\cite{extended_crossover_2d, monster, Spin_and_charge_2D, False_Mott_Hubbard_2d,  benchmark_leblanc,shift_1, benchmark_leblanc,shift_3}. In three dimensions, this crossover is replaced by a second order finite-temperature phase transition for all finite values of interactions~\cite{hirsch, ScalettarQMC3d, QMC, Khatami, Double_occupancy_entropy, sun2024boosting, DCA_size_scaling, DDMC, TUFRG, Connor, Renaud, Rohringer2011, Dual_fermions, Real_material_3D, tpsc3d, song2024extended, toschi-katanin-criticality}.

Away from half-filling, numerical results are much less readily available, as all known controlled numerical algorithms suffer from severe computational bottlenecks. Whilst there has been a lot of recent progress in unveiling the zero- and finite-temperature phase diagrams in two dimensions and the intricate relationships between pseudogap physics, stripe phases and d-wave superconductivity~\cite{xu_stripes_sdw, qin_absence, bo_xiao_finite_temperature, SimkovicPG, xu2024coexistence, scholle_unrestrictedHF}, much less is known for the three-dimensional doped counterpart. Here, the antiferromagnetic order has been observed to transition into magnetic spiral phases at the mean-field level of theory \cite{2nd_order_pert_theory, xu2013magnetic, scholle_unrestrictedHF} as well as with the more sophisticated algorithms, such as the dynamical vertex approximation (D$\Gamma$A)~\cite{Twist}, functional renormalization group (fRG)~\cite{TUFRG} and diagrammatic Monte Carlo~\cite{Connor}. Despite that, not much is known about the extent of the spiral phase in the phase diagram as well as about the dependence of the magnetic ordering vector $\mathbf{Q} = (q_x,q_y,q_z)$ on system properties, namely doping, temperature and interaction strength. 

The dynamical mean-field theory (DMFT) has over decades emerged as one of the work-horses of condensed matter physics and material simulation thanks to its comparatively low computational cost and versatility. In most  cases, DMFT is being used in a form that is restricted to solutions in the paramagnetic regime and this approach has notably led to the method's correct prediction of the metal-to-insulator transition in a number of materials. However, DMFT has recently been also increasingly successfully applied to the study of ordered phases, such as magnetic phases, e.g. in the square~\cite{GoremykinSpiralDMFT} and triangular~\cite{Wietek_triangular} lattices and the stripe (spin- and charge-) ordered states on a square lattice of the Hubbard model~\cite{peters_kawakami}. In three dimensions, DMFT has to date only been utilized in the paramagnetic regime~\cite{DeLeo2011DFMT, Twist}, yet it is the most natural setting for using it to study ordered phase and phase transitions.

In this work, we use a universal formalism of spiral DMFT which allows one to compute the leading magnetic $\mathbf{Q}$-vector from the evaluation of the free energy. We use this method to map out the entire magnetic phase diagram as a function of temperature, doping and interaction strength. We identify a number of different incommensurate spiral phases with ordering vectors of the  $(q,q,q)$, $(q,q,\pi)$ and $(q,\pi,\pi)$ kind. Importantly, we find a region where the competition between all incommensurate phases is extremely close and which exhibits negative compressibility, thus suggesting the possibility of spatial phase separation. Additionally, we provide an exhaustive set of results for thermodynamic quantities in the paramagnetic and ordered phases, including energy, magnetization, entropy and double-occupancy which are of direct relevance to cold atomic experiments. 

\emph{Model and method ---}
We are interested in the three-dimensional Hubbard model
\begin{equation}
    \mathcal{H} =
    \mathcal{H}_\mathrm{kin}+\mathcal{H}_\mathrm{int} =
    -t \sum_{ij, \sigma} c^\dagger_{i \sigma} c_{j \sigma}
    + U \sum_i n_{i \uparrow} n_{i \downarrow},
\end{equation}
where $c^\dagger_{i \sigma}$ creates a fermion with spin-$\sigma$ at the
position $\mathbf{R}_i$ of a cubic lattice with $N$ sites, $t$ is the nearest-neighbor hopping amplitude and $U$ is the onsite Coulomb repulsion. $\mathcal{H}_\mathrm{kin}$ and $\mathcal{H}_\mathrm{int}$ are respectively the kinetic and potential part of this Hamiltonian. We consider a co-planar spin spiral state in the $x-z$ plane described by a reciprocal vector $\mathbf{Q}$,
\begin{equation}
  \langle S^z_i \rangle + i \langle S^x_i \rangle = \frac{m}{2} \, e^{i \mathbf{Q} \cdot \mathbf{R}_i },
\end{equation}
where $m$ is the order parameter (the magnetization, $|m| \le 1$) and $S^\alpha = \sigma_\alpha / 2$ are the spin operators in the $\alpha$ direction. It is convenient to describe the spiral states in a rotating reference frame for the spins $\begin{pmatrix} \tilde{c}_{i \uparrow} & \tilde{c}_{i \downarrow} \end{pmatrix} = T^{-1}_{i} \begin{pmatrix}  c_{i \uparrow} & c_{i \downarrow} \end{pmatrix}$, where $T_i = \exp \left( -i \frac{\mathbf{Q} \cdot \mathbf{R}_i}{2} \sigma_y \right)$ is the rotation operator about the $y$ axis~\cite{fleck_spiral_dmft}. In this basis, and after Fourier transforming the kinetic term, the Hamiltonian takes the form
\begin{equation}
    \mathcal{H} =
    \sum_\mathbf{k}
    \begin{pmatrix}
    \tilde{c}_{\mathbf{k} \uparrow}^\dagger &
    \tilde{c}_{\mathbf{k} \downarrow}^\dagger
    \end{pmatrix}
    \begin{pmatrix}
    E_\mathbf{k} & i \, \eta_\mathbf{k} \\
    -i \, \eta_\mathbf{k} & E_\mathbf{k}
    \end{pmatrix}
    \begin{pmatrix}
    \tilde{c}_{\mathbf{k} \uparrow} \\
    \tilde{c}_{\mathbf{k} \downarrow}
    \end{pmatrix}
    + U \sum_i n_{i \uparrow} n_{i \downarrow},
\label{eq:H}
\end{equation}
where $E_\mathbf{k} = \frac{1}{2} \left( \epsilon_{\mathbf{k} +\mathbf{Q}} + \epsilon_{\mathbf{k} -\mathbf{Q}} \right)$, 
$\eta_\mathbf{k} = \frac{1}{2} \left( \epsilon_{\mathbf{k} +\mathbf{Q}} - \epsilon_{\mathbf{k} -\mathbf{Q}} \right)$ and $\epsilon_{\mathbf{k}}$ is the
dispersion of the cubic lattice.

We investigate the Hubbard model~\eqref{eq:H} using dynamical mean-field theory. In this approach, the lattice self-energy is approximated to be local $\Sigma_{\sigma \sigma'}^\mathrm{latt}(\mathbf{k}, i\omega_n) = \Sigma_\sigma^\mathrm{imp}(i \omega_n) \delta_{\sigma, \sigma'}$ and is identified with the self-energy of an auxiliary quantum impurity problem where a single correlated orbital is hybridized to an electronic bath. The structure of the bath is adjusted self-consistently in a way to ensure that the Green function of the correlated level $G^\mathrm{imp}(i\omega_n)$ equals the local lattice Green function within the DMFT approximation
\begin{equation}
  G^\mathrm{imp}(i\omega_n) = \frac{1}{N} \sum_{\mathbf{k}}
  \left\{ i\omega_n + \mu - \hat{\varepsilon}_\mathbf{k}
  -
  \begin{pmatrix}
      \Sigma^\mathrm{imp}_\uparrow & 0 \\
      0 & \Sigma^\mathrm{imp}_\downarrow
  \end{pmatrix}
  \right\}^{-1}
\end{equation}
where $\hat{\varepsilon}_\mathbf{k} = \begin{pmatrix} E_\mathbf{k} & i \, \eta_\mathbf{k} \\ -i \, \eta_\mathbf{k} & E_\mathbf{k} \end{pmatrix}$ and the chemical potential $\mu$ is adjusted in order to reach the target density.
We solve the quantum impurity problem using the CT-INT algorithm~\cite{rubtsov_jetp_2004, rubtsov_prb_2005} implemented with the TRIQS library~\cite{triqs_2015}. From the solution of the DMFT equations, we obtain the self-energy $\Sigma_\sigma^\mathrm{imp}(i\omega_n)$, the magnetization $m$, as well as other useful correlators such as $\langle \tilde{c}^\dagger_{\mathbf{k} \sigma} \tilde{c}_{\mathbf{k} \sigma'} \rangle$ and the local double occupancy $\langle n_{i \uparrow} n_{i \downarrow} \rangle$. As discussed below, this will allow us to determine useful thermodynamic quantities, in particular the entropy, the free energy and both the potential and kinetic energy.

\begin{figure}
\centering
\includegraphics[width=0.48\textwidth]{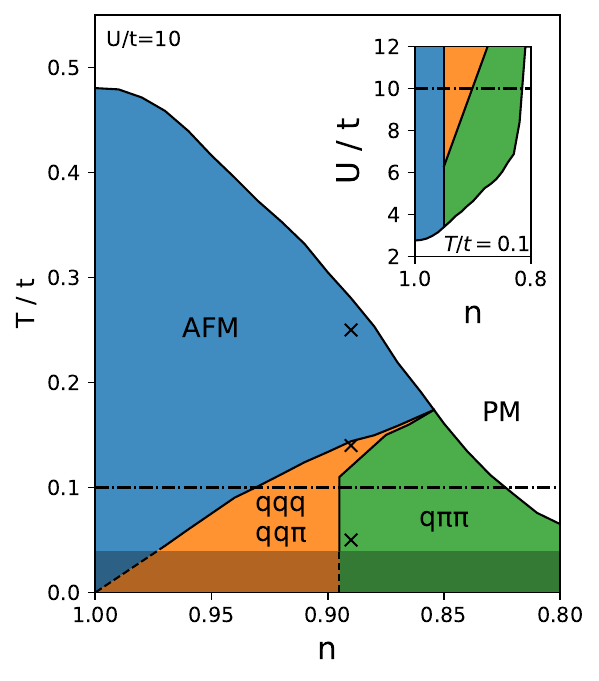}
\caption{DMFT phase diagram for spin-spiral orders in the cubic Hubbard model at $U/t = 10$. Below the lowest computed temperature $T/t=0.04$, we show linear extrapolations of the phases and their frontiers with dashed lines and shaded colors. Free energies shown in Fig.~\ref{fig:Q_optimization} are computed at the crosses.}
\label{fig:phase_diagram}
\end{figure}

\begin{figure}
\centering
\includegraphics[width=0.45\textwidth]{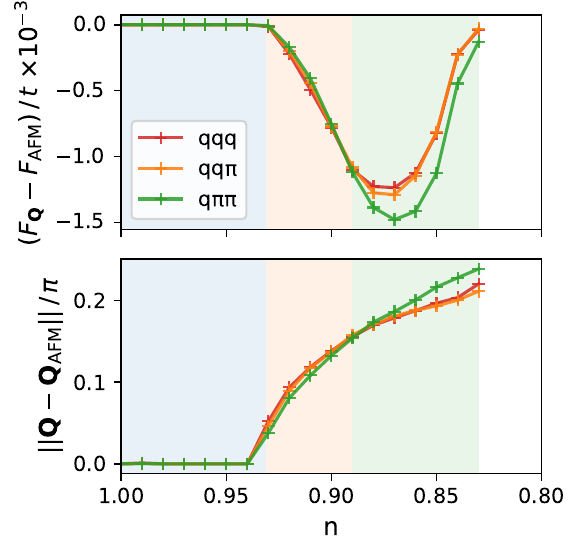}
\caption{Top panel: Free energy difference with AFM of the optimal ordering vectors of type $(q,q,q), (q,q,\pi)$ and $(q,\pi,\pi)$ at $T/t = 0.1$ (following the dash-dotted line of Fig.~\ref{fig:phase_diagram}). Bottom panel: Distance between the optimal ordering vector and $\mathbf{Q}_\mathrm{AFM}=(\pi,\pi,\pi)$.}
\label{fig:optimal_Q_and_F}
\end{figure}

\emph{Phase diagram ---}
In Fig.~\ref{fig:phase_diagram}, we present the phase diagram obtained within DMFT as a function of temperature and doping at a fixed interaction value of $U/t = 10$. At high temperature, the system is paramagnetic (PM). As temperature is decreased, magnetic order eventually sets at a Néel temperature which decreases with growing doping. The dominant magnetic order close to half-filling is antiferromagnetic (AFM) with wave vector $\mathbf{Q}_\mathrm{AFM} = (\pi,\pi,\pi)$. As the doping is increased, the system turns into an incommensurate spin spiral state (IC). The Néel transition is second order and the behavior of the magnetization close to the transition is compatible with the critical exponent $\beta = 1/2$ expected within DMFT, see Fig.~\ref{fig:magnetization_fit} in the End Matter.

In order to establish which incommensurate order is stabilized in the phase diagram, we have compared the free energies $F(\mathbf{Q})$ of different DMFT solutions, each computed at a fixed value of $\mathbf{Q}$. Obtaining very accurate free energies is in general a difficult task which requires the computation of the entropy from an integral over a broad set of solutions that start at a known limit, e.g. infinite temperature or vanishing occupation (see End Matter).
While this approach can provide useful qualitative information, we found that it is not accurate enough to distinguish different spiral states that are close in free energy. We, therefore, use a different strategy and compute the gradient of the free energy along typical paths in the Brillouin zone:
\begin{equation}
  \nabla_\mathbf{Q} F = \sum_\mathbf{k} \left\langle 
  {\tilde c}^\dagger_\mathbf{k} \,
  \nabla_\mathbf{Q}
  \hat{\varepsilon}_\mathbf{k}(\mathbf{Q}) \,
  {c}_\mathbf{k}
  \right\rangle.
  \label{eq:gradF}
\end{equation}
This quantity is a correlator that can be evaluated with a good precision. By integrating $\nabla_\mathbf{Q} F$ starting from $(\pi,\pi,\pi)$ in different directions, we obtain a precise estimate of the free energy difference between several solutions, which allowed us to determine which optimal $\mathbf{Q}$ vector is stabilized (more details are given in the End Matter).

In Fig.~\ref{fig:optimal_Q_and_F}, we show the free energy differences between the antiferromagnetic state and solutions with optimal ordering vectors $\mathbf{Q}$ taken along the directions $(q,q,q)$, $(q,q,\pi)$ or $(q,\pi,\pi)$. It appears that, with increasing doping, the incommensuration vector $\mathbf{Q}$ gradually moves away from $(\pi,\pi,\pi)$ and first takes values which are indistinguishable within our accuracy between vectors of type $(q,q,q)$ and $(q,q,\pi)$. This intermediate region has a broader extent for large values of $U$ (see inset of Fig.~\ref{fig:phase_diagram}). At larger doping $\sim 10\%$, the $\mathbf{Q}$ vector subsequently undergoes a transition to the $(q,\pi,\pi)$ vector. 

\begin{figure}
\centering
\includegraphics[width=0.45\textwidth]{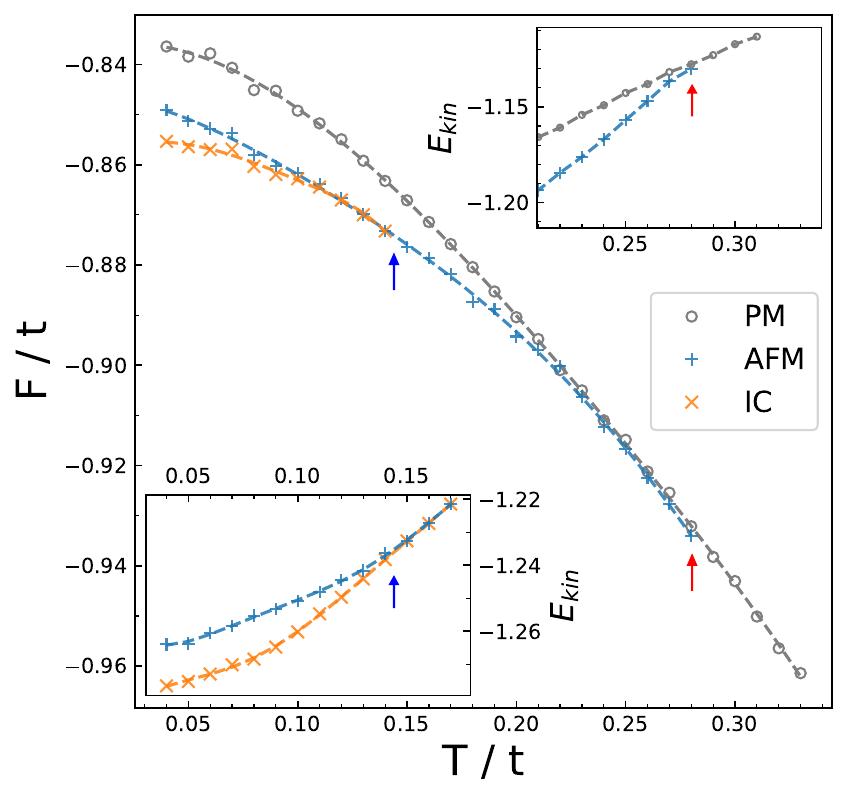}
\caption{Free energy of the PM, AFM and free-energy minimizing incommensurate spin spiral (IC) solution at $U/t = 10$ and $n=0.89$. The Néel transition takes place at $T/t=0.28$ (red arrow) and the AFM to incommensurate transition at $T/t = 0.14$ (blue arrow). Insets: Kinetic energy of the aforementioned states around the Néel (top) and incommensurate (bottom) transition temperatures.}
\label{fig:transition_mechanism}
\end{figure}

\emph{Nature of the transition ---}
We further investigate the nature of the transition between the different phases by following the evolution of the free energy and the kinetic energy along a temperature cut at a fixed density of $n=0.89$, see Fig.~\ref{fig:transition_mechanism}. It clearly appears that the transition from the paramagnetic state to the antiferromagnetic phase is kinetic energy driven (see top right inset). At the value of $U/t=10$ which we are considering, this is compatible with a Heisenberg mechanism for the Néel transition with the ordering reducing the effect of the Pauli exclusion when electrons hop to a neighboring site, thus favoring the kinetic term. At lower temperatures, the transition to an incommensurate state is also characterized by a kinetic energy gain (see bottom left inset). Indeed, the incommensuration of the magnetic order reduces the frustration induced by the doping of a perfect antiferromagnet.

\begin{figure}
\centering
\includegraphics[width=0.45\textwidth]{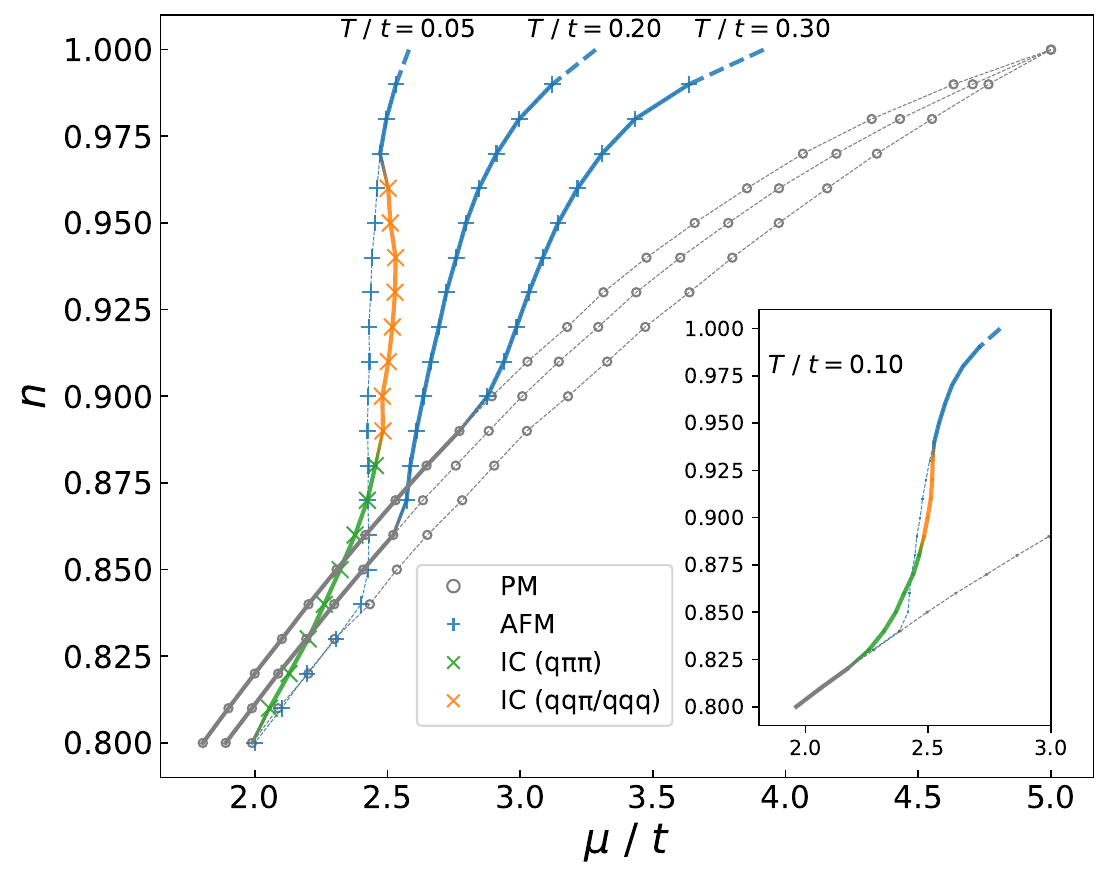}
\caption{Main plot: Density as a function of chemical potential for three different temperatures and for the PM,
AFM and incommensurate spin spiral (IC) solutions. The solid lines represent stable solutions, dashed ones are higher-energy underlying phases. Inset: Same plot at $T/t=0.1$, separated for readability.}
\label{fig:n_vs_mu}
\end{figure}

\emph{Thermodynamic stability ---}
In Fig.~\ref{fig:n_vs_mu}, we show the behavior of the density as a function of chemical potential and temperature for the different phases. A first observation is that the paramagnetic solution remains metallic at half-filling for $U/t=10$, even at the lowest temperature $T/t=0.05$. The onset of magnetism instead drives the system to be insulating with a clear gap at temperatures below $T/t = 0.3$. In the regime of temperatures where the AFM solution minimizes the free energy, i.e. $T/t \gtrsim 0.2$, the system has a positive compressibility $\partial n / \partial \mu > 0$ which increases with decreasing temperature. At low temperature, $T/t = 0.05$, the compressibility of the AFM solution becomes almost infinite, a behavior which is compatible with the stabilization of a more favorable incommensurate state. It is interesting to observe that the incommensurate solution has a compressibility that gradually increases up to densities $n \simeq 0.93$. At that value, the compressibility diverges and even becomes negative for densities between $0.93 \lesssim n \lesssim 0.97$. This suggests that the system would favor a phase separation in the incommensurate regime close to the AFM phase, hinting at a low-temperature first-order phase transition that can only be indirectly observed within our method. This may also be indicating that another order, not captured within our approach, such as a stripe order, would be stabilized in this regime of parameters, as has been observed at the mean-field level in Ref.~\cite{scholle_unrestrictedHF}. 

\begin{figure}
\centering
\includegraphics[width=0.48\textwidth]{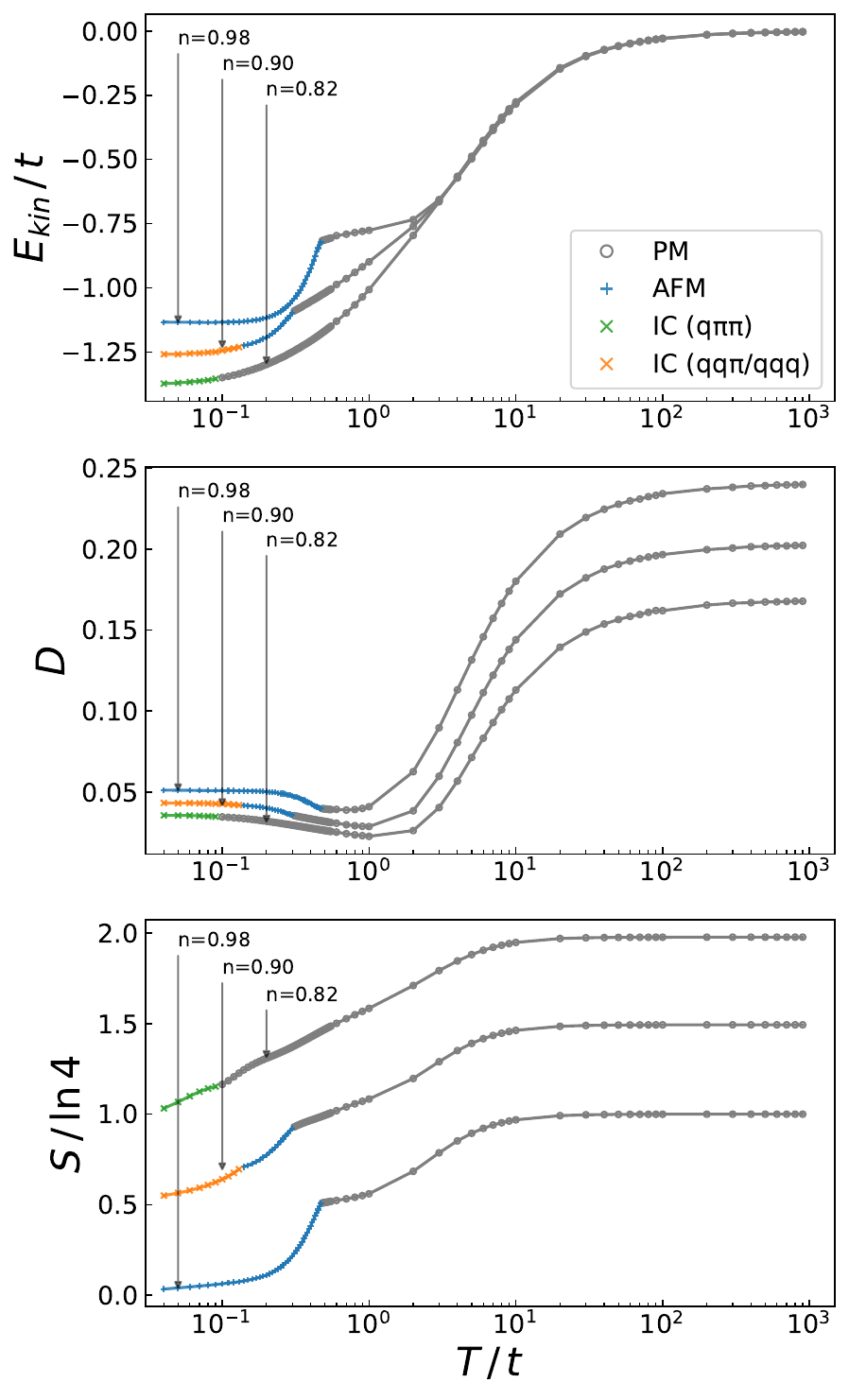}
\caption{Temperature cuts for thermodynamically relevant parameters: Kinetic-energy per site (top), double-occupancy (middle), entropy per site (bottom) shifted by $\ln2$ for $n=0.9$ and $\ln4$ for $n=0.82$. }
\label{fig:thermodynamics}
\end{figure}

\emph{Other thermodynamic quantities ---}
The kinetic energy $E_\mathrm{kin}=\langle\mathcal{H}_\mathrm{kin}\rangle$, double occupancy $D=\langle n_{i \uparrow} n_{i \downarrow} \rangle$ (which, up to a factor $U$, is identical to the potential energy $\langle \mathcal{H}_\mathrm{int} \rangle$) and entropy are displayed as a function of temperature for three different densities in Fig.~\ref{fig:thermodynamics}. There are several distinct temperature regimes that can be identified. For high temperatures $T/t \gtrsim 10$, interaction effects have a very weak effect on the kinetic energy and entropy, which we have checked are both very close to those of free electrons. Only the double occupancy starts to decrease in this regime. In the range $2 \lesssim T/t \lesssim 10$ the interaction effects start to set in: The kinetic energy is larger than for a non-interacting system, while still very weakly density dependent, and the entropy decreases as for weakly interacting electrons. The situation changes at $T/t \simeq 2$ where an inflection point is seen in the double occupancy. Below that temperature and before entering the ordered phase, the electrons gain coherence while experiencing strong correlations. As a result, a Pomeranchuk regime is observed: An increase of the interaction $U$ leads to more localization and a corresponding increase of the entropy. The relation $\partial D / \partial T = - \partial S / \partial U$ shows that the Pomeranchuk effect induces a range of temperatures where the double occupancy increases with decreasing temperature, especially at larger doping where the transition temperature is lower. Eventually, the systems undergoes a magnetic phase transition, which, as discussed above, is driven by a kinetic energy gain. The construction of the magnetic state comes with a very rapid drop in the entropy as $T$ goes to zero.

\emph{Conclusions ---}
In this work, we have systematically studied the phase diagram of the three dimensional Hubbard model within a DMFT approach, which considers all possible spin spiral ordering vectors. At low temperature and close to half-filling, the commensurate antiferromagnetic state is favored. But as doping is increased, a spiral state appears, with an incommensurate ordering vector either along the $(q,q,q)$, $(q,q,\pi)$ or $(q,\pi,\pi)$ direction. At intermediate to large interaction strength, the transition to the magnetic state is kinetic-energy driven and preceded by a Pomeranchuk regime, where the strongly correlated metal sees an increase in the double occupancy with decreasing temperature.
At very low temperature, an investigation of the compressibility hints towards the possibility of an unstable spiral state or a phase separation. It would be interesting to extend the current study to inhomogeneous phases using iDMFT~\cite{peters_kawakami} and, in particular, study whether a stripe phase would be stabilized, as discussed e.g. within static mean-field theory~\cite{scholle_unrestrictedHF}.

\emph{Acknowledgments ---}
The authors thank  Renaud Garioud, Antoine Georges and Thomas Schäfer for fruitful discussions. This work was granted access to the HPC resources of TGCC and IDRIS under the allocations A0150510609 attributed by GENCI (Grand Equipement National de Calcul Intensif). It has also used high performance computing resources of IDCS (Infrastructure, Données, Calcul Scientifique) under the allocation CPHT 2024. This work has been supported by the Simons Foundation within the Many Electron Collaboration framework.

\appendix

\section{Paramagnetic to spiral ordering phase transition}

\begin{figure}
\centering
\includegraphics[width=0.48\textwidth]{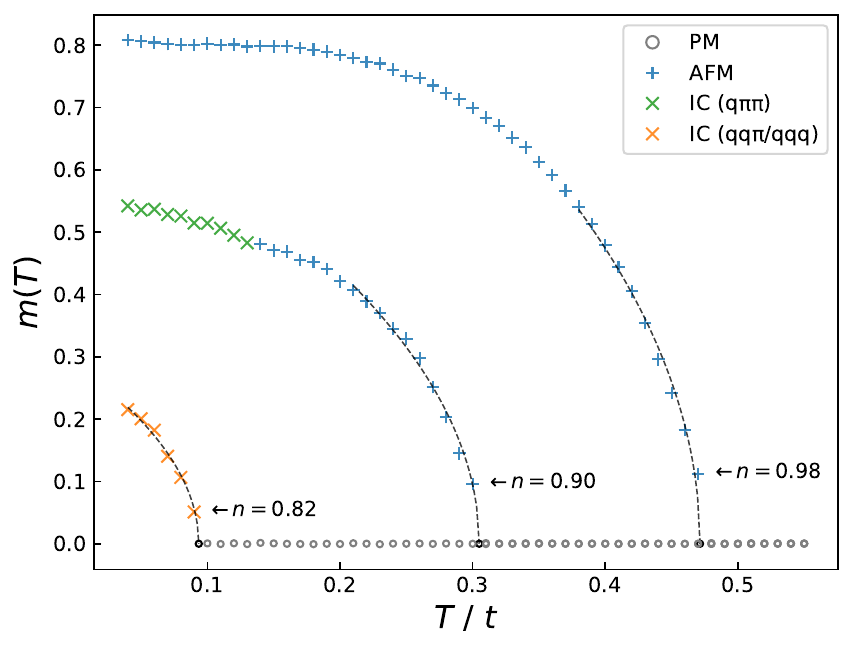}
\caption{Magnetization curves at various fillings. The dashed lines correspond to a fit with critical exponent $\beta=1/2$.}
\label{fig:magnetization_fit}
\end{figure}

In order to accurately determine the Néel temperature, we fit the magnetization versus temperature curve in the rotating reference frame. The best fits are obtained with the expected critical exponent $\beta = 1/2$
\begin{equation}
    \begin{cases}
        m \propto (T_\mathrm{Neel}-T)^\beta &(T < T_\mathrm{Neel})\\
        m = 0 &(T > T_\mathrm{Neel})
    \end{cases}
\end{equation}
Example results in Fig.~\ref{fig:magnetization_fit} show excellent agreement with the data.

\section{Determination of the optimal spiral ordering vector}

\begin{figure}
\centering
\includegraphics[width=0.48\textwidth]{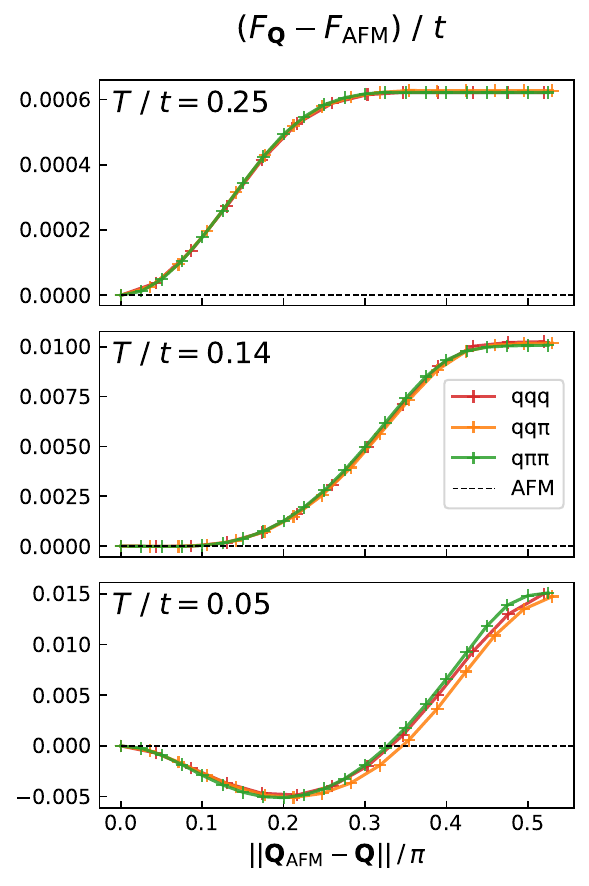}
\caption{Free energy difference between AFM and incommensurate states at filling $n=0.89$, interaction $U/t = 10$ at temperatures corresponding to crosses found in Fig.~\ref{fig:phase_diagram}. The AFM ordering vector changes from being a global minimum to a local maximum at the incommensurate transition.}
\label{fig:Q_optimization}
\end{figure}

In order to find the optimal ordering vector for spin spiral solutions, we integrate Eq.~\eqref{eq:gradF} along high-symmetry paths in the Brillouin zone: $(q,q,q)$, $(q,q,\pi)$ or $(q,\pi,\pi)$ with $q \in [0, \pi]$. As result, we obtain the free energy difference between the antiferromagnetic state and solutions with given ordering vectors $\mathbf{Q}$. Three typical examples are shown in Fig.~\ref{fig:Q_optimization} for a fixed density $n=0.89$ and three temperatures. At $T/t = 0.25$ (upper panel), the antiferromagnetic solution minimizes the free energy and appears as a global minimum. As temperature is decreased, this global minimum turns into a local maximum, where the optimal solution corresponds to an incommensurate spiral state. Note that $F(\mathbf Q_\mathrm{AFM})$ is always an extremum by symmetry of the Brillouin zone. The temperature $T_\mathrm{spiral}$ where the transition to an incommensurate solution occurs can be accurately determined by polynomial fitting of the Hessian and solving for roots. At low temperature (bottom panel), the optimal ordering vector can be determined by finding the minimum of the free energy difference.

\section{Entropy calculation}

We compute the entropy of the paramagnetic and magnetic solutions by integration of the fundamental thermodynamic relation \begin{equation} \mathrm{d}E = T\mathrm{d}S \Rightarrow \frac{\partial S}{\partial T}  = \frac{1}{T}\frac{\partial E}{\partial T}
\end{equation} with the entropy per site of the Hubbard model at infinite temperature and filling $n$ 
\begin{equation}
     S_\infty = \ln(4) - n\ln(n) - (2-n)\ln(2-n)
\end{equation} as a boundary condition. Above the Néel temperature, we fit the internal energy $E$ by a polynomial in $\ln T$ to account for the scattering of data points. In the ordered phase, we use a polynomial in $T$, constrained to impose continuity of the internal energy and the specific heat $\partial E / \partial T$ at $T_\mathrm{Neel}$ (consistent with a second order phase transition) and vanishing entropy at zero temperature.

\end{document}